\documentclass[10pt,twocolumn]{article}
\usepackage{times}
\usepackage[top=22mm,left=25mm,right=25mm,bottom=26mm,columnsep=6mm]{geometry}
\usepackage[english]{babel}
\usepackage[T1]{fontenc}

\usepackage{graphicx}
\usepackage{textcomp}
\usepackage[english]{babel}
\usepackage{hyperref}
\usepackage{url}
\usepackage{balance}
\usepackage{adjustbox}
\usepackage[utf8]{inputenc}

\usepackage[small,compact]{titlesec}
\titleformat{\subsubsection}[runin]
            {\normalfont\it}
            {\thesubsubsection}{0.5em}{}[:]

\title{Industrial Requirements for Supporting \\AI-Enhanced Model-Driven Engineering}

\usepackage{authblk}
\author[1]{Johan Bergelin}
\author[2]{Per Erik Strandberg}

\affil[1]{Mälardalen University, Västerås, Sweden}
\affil[2]{Westermo Network Technologies AB, Västerås, Sweden}

\date{\small{Accepted to the 4'th Workshop on Artificial Intelligence and Model-driven Engineering
(MDE Intelligence), 2022}}

\begin{document}
\maketitle

\section*{Abstract}

There is an increasing interest in research on the combination of AI techniques and methods with MDE. 
However, there is a gap between AI and MDE practices, as well as between researchers and practitioners.
This paper tackles this gap by reporting on industrial requirements in this field.
In the AIDOaRt research project, practitioners and researchers collaborate on \textit{AI-augmented automation} supporting modeling, coding, testing, monitoring, and continuous development in cyber-physical systems. The project specifically lies at the intersection of industry and academia collaboration with several industrial use cases.
Through a process of elicitation and refinement, 78 \textit{high-level requirements} were defined, and generalized into 30 \textit{generic requirements} by the AIDOaRt partners.
The main contribution of this paper is the set of generic requirements from the project for enhancing the development of cyber-physical systems with artificial intelligence, DevOps, and model-driven engineering, identifying the hot spots of industry needs in the interactions of MDE and AI.
Future work will refine, implement and evaluate solutions toward these requirements in industry contexts.

\noindent
\textbf{Keywords:} 
Model-driven engineering, Artificial intelligence, Requirements, Cyber-physical systems

\section{Introduction}
Model-driven engineering (MDE) has been adopted and applied in industry for quite some time  \cite{schmidt2006model}. Recently, a paper by Bucchiarone et al.\ discusses the future of modeling \cite{bucchiarone2021future}. In their paper it is identified that a large area of potential for MDE improvement and maturity is Artificial intelligence (AI). They argue that applications such as recommender systems or modeling assistance utilizing AI contains considerable potential. Similarly Cabot et al.\ \cite{cabot2017cognifying} identify the potential improvements of implementing AI for modeling, reasoning that even a small amount of intelligence integrated into an existing process can boost the effectiveness by large amounts. However as noted by Cabot et al. \cite{cabot2017cognifying}, it is not trivial to implement MDE with integrated AI and several hurdles need to be overcome, and perhaps the largest issue for technical advances is the lack of available high-quality training data.

However, the literature is limited in reporting on the needs and challenges from an industry and practitioner perspective. Existing works mostly regard academic contributions or consider fundamental techniques or principles, often extending already mature technology. Garousi et al.\ performed a literature review on the challenges in industry and academia collaboration \cite{garousi2016challenges}, and identified several collaboration challenges, e.g.\  
results in academia that are not applicable to industry, different interests and objectives, different perceptions on what solutions and outcomes are useful, etc. 
They also 
provide a set of best practices, of which the most important one is 
to base research on real-world problems. 

\begin{figure}[bt]
\begin{center}
\includegraphics[width=\linewidth]{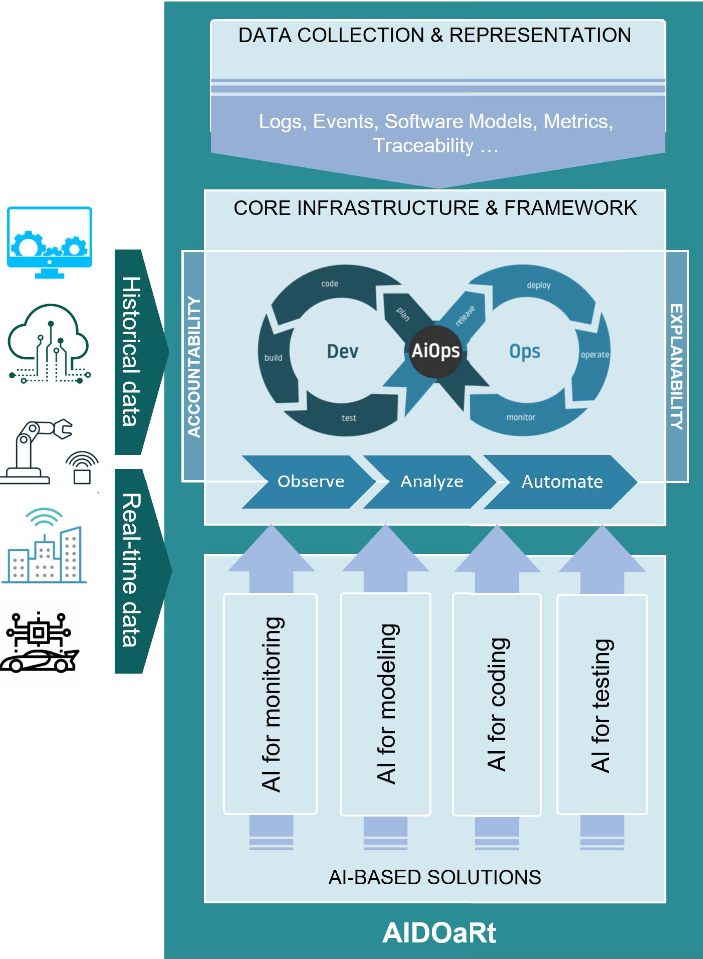}
\caption{
Conceptual architecture of AIDOaRt from \cite{eramo2021aidoart}.
\label{fig:AIDOaRtVennDiagram}
}
\end{center}
\end{figure}

\begin{figure*}[th!]
\begin{center}
\includegraphics[width=0.99\linewidth]{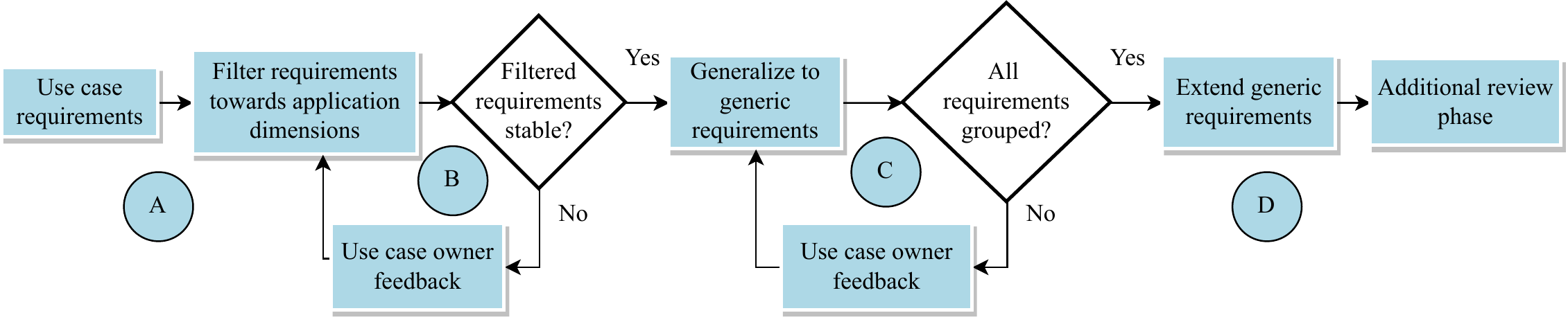}
\caption{A flow chart for the requirements elicitation and refinement process.
\label{fig:AidoartFlowChart}
}
\end{center}
\end{figure*}

AIDOaRt (AI-augmented automation supporting modelling, coding, testing, monitoring and continuous development in cyber-physical systems, \href{https://www.aidoart.eu/}{aidoart.eu}) is a European research project where many partners define, implement and evaluate solutions for AI-powered software development processes for cyber-physical systems (CPS). In particular the project aims to find potential solutions and beneficial interactions between the three domains of AI, MDE, and DevOps. The conceptual architecture is vizualized in Figure~\ref{fig:AIDOaRtVennDiagram}. The project strives to provide a framework containing a set of various capabilities tackling identified industrial challenges, specifically targeting solutions that combine two or more of the central domains of AI, MDE and DevOps. Partly the project will also report on the measured impact of these solutions in industrial contexts.

In the context of AIDOaRt, several high-level requirements have been extracted from industrial case studies by several companies, as well as a set of generic requirements that can be thought of as an abstraction, aiming to capture common aspects and challenges. The central contribution of this paper is the set of generic requirements from the industrial partners in the AIDOaRt project (see Table~\ref{tab:generic-requirements}). We are interested in discussing the implications of these requirements for a sub-set of the AIDOaRt project, specifically the intersection of AI and MDE identified as the most attractive combination of the key domains during the requirements elicitation activities.

\section{Background}
\label{aidoart-generic-requirements-background}

An AI system can be described as an autonomous decision-maker striving for some goal in an environment and often includes a learning (or Machine Learning, ML) aspect such that the system improves its decision-making when having more data
\cite{legghutter2007collection}.
MDE can be described as an approach where a model is the primary artifact of the development process \cite{brambilla2017model}, often with automatic code generation, which encompasses the other activities within software engineering (SE) \cite{akdur2018survey}.
Also continuous practices such as Dev\-Ops can be described as a set of practices used by organizations while striving to develop, deploy and get feedback in fast cycles \cite{shahin2017continuous}.
Furthermore, with CPS, software engineering practices are given a unique set of challenges in that these systems both have hardware and software and often act in an environment they interact with \cite{garousi2018testing}.

In recent years, much research has come out of the intersections of AI, MDE and DevOps, e.g.\ Combemale and Wimmer present a vision for model-based DevOps practices \cite{combemale2019towards}.
Further interest in applications such as presented in \cite{ries2021mde} is rising, where MDE aims to improve deep learning data set requirements using UML. From the literature it seems that MDE and AI can have positive interactions to tackle several observed challenges in the industrial setting, and as previously mentioned AI is viewed as a fundamental enabler for future MDE maturity \cite{bucchiarone2021future}. The AIDOaRt project and its related activities in part aim to deliver such solutions to investigate and evaluate the applicability and results in industrial settings.

AIDOaRt is a three year EU (ECSEL-JU) project with more than 30 partners and more than 3000 person months,
aiming at improving the continuous development of CPS with a wide range of use cases in domains such as railway, automotive, restaurants, etc. By combining principles from AI, DevOps and MDE, a model-based framework is developed by the partners. The project strives to enhance the DevOps and software development life-cycle by including AI and ML techniques, as well as support monitoring of both historic and real-time data. 
A set of deliverables are used throughout the project to document and communicate results. The topic of this paper overlaps with the first deliverable D1.1: ``Use cases and requirements specification.'' Since the deliverable is unavailable for the public audience, we limit our reporting to the generic requirements for legal and ethical reasons, abstracting partner-specific details.

\section{Eliciting Generic Industry Requirements}
\label{aidoart-generic-requirements-method}

The AIDOaRt project is evaluated via 15 industrial case studies, in particular towards a set of requirements refined from original input from the partners. The elicitation process is illustrated in Figure~\ref{fig:AidoartFlowChart}.
First (See A in Figure~\ref{fig:AidoartFlowChart}), the industrial use case providers defined 78 high-level requirements.
The high-level requirements had a unique identifier, a definition in natural language, a type (functional, performance, usability, reliability, security, maintainability or portability), priority (five-graded scale from lowest to highest), links to application domains (MDE, AI, DevOps), a rationale in natural language, comments in natural language, the current fulfillment level of the requirement for the partner, as well as the date for when the requirement is to be fulfilled.
The initial high-level requirements were use case dependant, extracted via processes internal to each use case provider, and coupled with that particular partner. 
Next (B), internal filtering of the requirements grouped them into the five application dimensions of AIDOaRt: \textit{Requirements Engineering} (RE), \textit{Modeling} (Mod.), \textit{Coding} (Code), \textit{Testing} (Test), and \textit{Monitoring} (Mon.).
Then (C), the filtered requirements were analyzed for logical and semantical coherence between partners, and some were further refined to align the common understanding.
Once the requirements were stable and filtered, a working group for each application dimension was created based on interest from the project partners. 
The working groups generalized the filtered requirements in its dimension. Feedback from the use case owners provided clarity and refinement on requirements not matched to any generic requirement until all requirements had a link to at least one generic requirement.
The working groups then extended the initial generic requirements. 
Finally, an additional review phase confirmed the consistency and overall quality resulting in 30 generic requirements (D). 
Two generic requirements had sub-requirements, these have been somewhat simplified in this paper, and one requirement (Mod.08) was created in the Mod working group, and kept among the generic requirements for its importance, despite not being linked to any of the high-level requirements directly.

To manage this process the project utilized model-based requirements engineering (MBRE) services with Modelio SaaS. Prior experiences with this tool in EU projects have been positive \cite{sadovykh2021applying}.
From the MBRE activities traceability links and relations were defined between the 78 high-level requirements (not the generic requirements) and project dimensions. Figure~\ref{fig:RequirementRelations} relates the requirements related to AI/ML to the five application dimensions (RE: Requirements Engineering, Mod: Modeling, Code: Coding, Test: Testing and Mon: Monitoring). The figure aims to illustrate the hot spots in the project regarding AI, visualized with circle sizes, and it is clear that AI/ML in combination with Modeling is of interest to the industrial partners. 
The 30 industrial generic requirements 
are presented in Table~\ref{tab:generic-requirements}, where they are grouped by application dimension. The number of specific requirements linked to the generic requirements is indicated in the N column. E.g., the most relevant requirement (in terms of instantiations) is Mod.04: ``The system shall verify models with semi-automatic model synthesis.'' It is worth noting that a high-level requirement can be linked to one or more generic requirements, therefore the sum of all links (N) is greater than the sum of the high-level requirements (78).

\section{Discussion}
\label{aidoart-generic-requirements-discuss}

In this section, we discuss the implications of the generic requirements. We focus on research on modeling and implications for practitioners with respect to AI/ML.

Compared to previous work on the intersection of AI and MDE \cite{battina2016ai, mussbacher2020towards}, we present observations from industrial settings. Battina \cite{battina2016ai} describes a vision for how a model-driven framework can incorporate AI and DevOps practices to meet challenges observed in CPS development. Mussbacher et al.\ \cite{mussbacher2020towards} provide a framework for modeling assistance utilizing MDE, where they observe a need for such assistance due to the increasing complexity in modeling activities, reflected in \textit{Mod.04} and \textit{Mod.06}. From the literature, it is clear that the authors identify a need or potential for the collaboration of MDE and AI techniques and present potential means of addressing these aspects.

MDE, as mentioned by Bucchiarone et al.~\cite{bucchiarone2021future}, has seen a lot of success and progress to this point. They mention that MDE improves both informal modeling and low-code development, increasing collaboration and inclusion of non-experts. Further, they identify that model-based systems engineering (MBSE) has aided significantly in the engineering practices of developing CPS systems, specifically reliable systems. Further, they argue that AI is among the most important open challenges to address when it comes to MDE, stating that it cannot be denied that its advances in recent years have drastically changed many processes in software. Introducing AI with modeling methods and tools is a topic that recently has started to show promising results, however the real-world application is still lacking and is reflected in the requirements in Table \ref{tab:generic-requirements}. 

Our set of requirements has a strong industry focus, and represents requirements created mainly through industry practitioners in the context of CPS, reporting the needs directly from the industry, a challenge often observed from industry-academia collaboration \cite{garousi2016challenges}. A subset of our requirements also explicitly targets the domain of AI and MDE, and as far as we can tell, no previous publication mentions practitioners' needs in terms of industry requirements. Thanks to the layer or generalization (and anonymization), we can share requirements that otherwise would have remained hidden under corporate secrecy.

For academia these requirements could be used as a ``sanity check'' to position their solutions in terms of reported industry needs. Further it could open for discussions and workshops between academia and industry, tied as the most observed best practice along with basing research on real-world problems, in the review by Garousi et al.~\cite{garousi2016}. A successful outcome from the reported requirements is increased collaboration with industry and academia towards the same goals via a shared vision.

\begin{figure}[bt]
\begin{center}
\includegraphics[width=\linewidth]{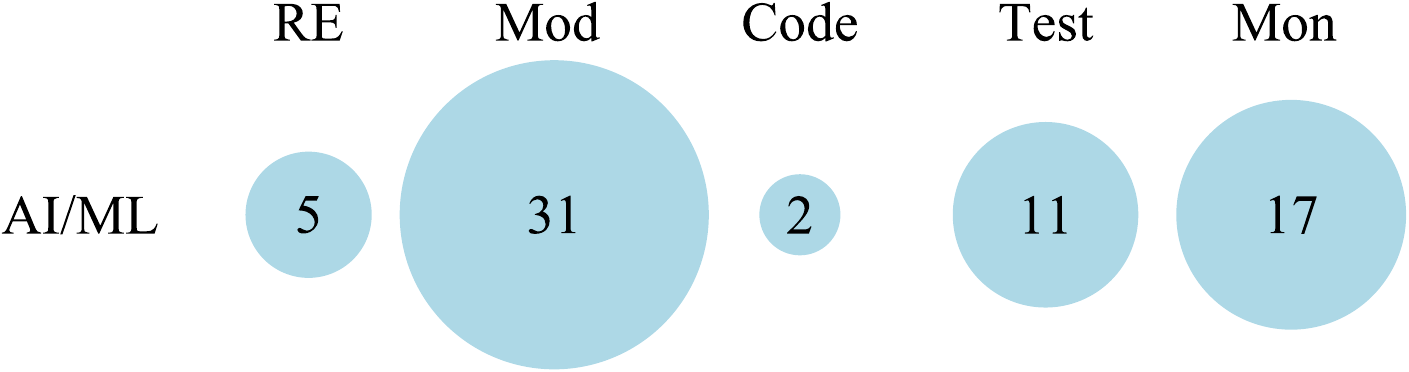}
\caption{Relationship between the 66 AI/ML related (out of 78 total) high level requirements towards the five application dimensions.
\label{fig:RequirementRelations}
}
\end{center}
\end{figure}

Figure~\ref{fig:RequirementRelations} and Table~\ref{tab:generic-requirements} (in particular requirements annotated with \textit{Mod}),
seems to indicate that industry practitioners see AI as a candidate approach for improving the modeling of CPS, consistent with the academic literature. Further the requirements highlight the need for solutions dealing with models at higher levels of abstraction, often regarding formulation or verification of models. Indeed it seems that from industry \textit{AI for modeling} is of particular interest, and the more significant requirements (Mod.01, Mod.02, Mod.04, Mod.06) all involve AI techniques or methods to some extent. In addition to utilizing AI, there is an interest in involving automation in the processes. We observe that the requirements align with previously reported AI opportunities for modeling \cite{cabot2017cognifying, bucchiarone2021future}.
E.g., Bucchiarone et al. \cite{bucchiarone2021future} suggest that modeling bots could assist in identifying issues or give advice to modelers. The generic requirements could be interpreted as target areas for such bots in industry. 

Apart from the implication of MDE and AI the requirements presented in Table~\ref{tab:generic-requirements} provide additional insights into industrial needs in the development of CPS. Requirements related to requirements engineering (\textit{RE} in Table~\ref{tab:generic-requirements}) seem to indicate the need for interpretation of semi-structured language along with suggestions for practitioners and consistency verification. Requirements related to coding practices (\textit{Code} in Table~\ref{tab:generic-requirements}) seem to foresee benefits from AI methods to both refine and abstract specifications and implementations respectively. Requirements for testing (\textit{Test} in Table~\ref{tab:generic-requirements}) in a similar fashion as Modeling seems to be centered around verification and specifically test case generation for models via AI techniques, specifically on high-level models. The monitoring requirements (\textit{Mon} in Table~\ref{tab:generic-requirements}) are the most numerous and seem to indicate a wide set of needs. The more prevalent requirements all point to a need to improve the coverage of data collection (\textit{Mon.1.1 and } \textit{Mon.1.2}), and the largest requirement not related to data collection (Mon.2.8) regards predictions of future behavior based on observations, presumably via AI methods and techniques which fits well with the highlighted need of data collection.

Overall, ethical aspects of applied AI are a hot topic, and we partially cover it in previous work \cite{strandberg2021ethical}.
Jobin et al.~\cite{jobin2019global} found that transparency and justice/fairness were top ethical principles in guidelines for ethical AI \cite{jobin2019global}.
The AIDOaRt project strives for AI approaches that are beneficial, safe and responsible \cite{eramo2021aidoart}. Furthermore, the concepts of accountability and explainability are highlighted in the conceptual architecture of AIDOaRt (in Figure~\ref{fig:AIDOaRtVennDiagram}). However, only one generic requirement partially covers an ethical challenge in that an AI ought to be able to motivate its actions (\textit{Mon.6}). This falls under the most prevalent principle of transparency in the classification by Jobin et al. Future research could explore how, and how well, AIDOaRt complied with ethical principles.

\section{Conclusion and Future Work} \label{Sec:Conclusions}

In this paper, we report on the industry needs and hot spots for integrating AI techniques and methods for MDE. Previous work has identified that MDE processes could be improved or integrated with AI to meet specific challenges in the increasing maturity and complexity of MDE. We present our findings via a set of generic requirements extracted from industrial partners via the AIDOaRt project, and discuss what implications this has on the research. Specifically, we identify that industry has a need for utilizing AI to improve verification activities, preferably utilizing some level of automation. 

Future work could explore how well these requirements are covered in previous work, explore potential gaps in them, and if they can support work by other groups of practitioners or researchers. Additionally, the AIDOaRt project will develop and evaluate solutions for the identified challenges in industrial settings, further aiming to bridge the gap between industry and academia via real-world challenges.

\section*{Acknowledgments}
This work was funded by the AIDOaRt project, an ECSEL
Joint Undertaking (JU) under grant agreement No. 101007350.

\noindent
\textbf{Author Contributions}\phantom{i}
Methodology and Investigation: JB, PES, and AIDOaRt partners.
Writing and Reviewing: JB, and PES.
Supervision: PES.

\bibliographystyle{abbrv}
\bibliography{references}

\begin{table*}[p]
\begin{center}
  \caption{Generic requirements extracted from the 78 high-level requirements in the AIDOaRt project covering AI, DevOps, and MDE. 
  The requirements have been slightly rephrased for anonymization and generalization.
  The second column, N, relates to the number of high-level requirements related to each generic requirement.
  }
  \label{tab:generic-requirements}
  \begin{adjustbox}{width=0.99\textwidth}
  \begin{tabular}{lcp{13.8cm}}
    \hline
    Id       &  N  & Requirement \\
    \hline
    RE.1     & 2 & The system shall translate requirements from semi-structured language to formal language.  \\ 
    RE.2     & 1 & The system shall refine requirements expressed in formal language from analysis results. \\
    RE.3     & 2 & The system shall produce suggestions or prescriptions for the requirements or systems engineer. \\ 
    RE.4     & 2 & The system shall verify consistency of requirements from analysis results. \\ 
    \hline
    Mod.01   & 6 & The system shall verify high-level models with AI techniques. \\
    Mod.02   & 5 & The system shall generate tests using formal models, automated reasoning and/or AI methods. \\
    Mod.03   & 2 & The system shall identify system failures using AI-based models. \\
    Mod.04   & 8 & The system shall verify models with semi-automatic model synthesis.  \\
    Mod.05   & 3 & The system shall abstract models with AI-based approaches. \\
    Mod.06   & 6 & The system shall provide easy configuration via AI-based methods. \\
    Mod.07   & 2 & The system shall represent AI with a Model-based approach. \\
    Mod.08   & - & The system shall perform retrospective analysis of model quality using AI-augmented methods. \\
    Mod.09   & 1 & The system shall extend standards modeling languages to support aspect-oriented system development. \\
    Mod.10   & 3 & The system shall deploy a solution on the cloud as a set of corresponding services with execution traceability. \\
    Mod.11   & 1 & The system shall integrate DevOps workflows with CI/configuration models.  \\
    \hline
    Code.1   & 1 & The system shall write abstractions of implementation with AI methods.  \\
    Code.2   & 1 & The system shall write refinement of specification with AI methods.  \\
    Code.3   & 1 & The system shall import/export variables/procedures between code and models. \\
    \hline
    Test.1   & 7 & The system shall generate test cases for high-level models with AI techniques. \\
    Test.2   & 1 & The system shall perform automatic execution of test cases.  \\
    Test.3   & 4 & The system shall perform automatic verification of system architectures.  \\
    Test.4   & 2 & The system shall perform automated evaluation of testing results.  \\
    Test.5   & 2 & The system shall report or fix the problems detected in the testing phase.  \\
    Test.6   & 1 & The system shall integrate and analyze the testing phase into the DevOps pipeline.  \\
    \hline
    Mon.1    & - & The system shall access different artifacts.  \\
    \phantom{x}Mon.1.1  & 7 & \phantom{x}The system shall access off-line data. \\
    \phantom{x}Mon.1.2  & 7 & \phantom{x}The system shall access on-line data.  \\
    \phantom{x}Mon.1.3  & 4 & \phantom{x}The system shall access system logs and execution traces. \\
    \phantom{x}Mon.1.4  & 3 & \phantom{x}The system shall access system states such as resource usages during monitoring.  \\
    \phantom{x}Mon.1.5  & 2 & \phantom{x}The system shall access traffic between nodes. \\
    \phantom{x}Mon.1.6  & 1 & \phantom{x}The system shall access hardware input/output. \\
    \phantom{x}Mon.1.7  & 1 & \phantom{x}The system shall access Documents expressed in natural language. \\
    Mon.2    & - & The system shall support monitoring for many purposes.  \\
    \phantom{x}Mon.2.1  & 2 & \phantom{x}The system shall monitor with the purpose of identifying violations to pre-defined requirements.  \\
    \phantom{x}Mon.2.2  & 3 & \phantom{x}The system shall monitor for the purpose of identifying deviations, anomalies or security events. \\
    \phantom{x}Mon.2.3  & 3 & \phantom{x}The system shall monitor for the purpose of identifying clusters of anomalies of some sense. \\
    \phantom{x}Mon.2.4  & 2 & \phantom{x}The system shall monitor for the purpose of identifying (unwanted) trends or drifts.  \\
    \phantom{x}Mon.2.5  & 3 & \phantom{x}The system shall monitor for the purpose of identifying deviations between a system and a model.  \\
    \phantom{x}Mon.2.6  & 1 & \phantom{x}The system shall monitor for the purpose of identifying root-causes to anomalies/deviations. \\
    \phantom{x}Mon.2.7  & 3 & \phantom{x}The system shall monitor for the purpose of minimizing downtime or other inefficiencies/waste.  \\
    \phantom{x}Mon.2.8  & 5 & \phantom{x}The system shall provide predictions of future (unwanted) behaviours based on observed features. \\
    \phantom{x}Mon.2.9  & 1 & \phantom{x}The system shall monitor for the purpose of tracking system usage or user activities. \\
    Mon.3    & 1 & The system shall discriminate between levels of automation during monitoring.  \\
    Mon.4    & 1 & The system shall produce artifacts 
    for re-use and further training. \\
    Mon.5    & 1 & The system shall provide human-in-the-loop functionality supporting heterogeneous data sources.  \\
    Mon.6    & 1 & The system shall motivate its actions during monitoring. \\
    \hline
  \end{tabular}
  \end{adjustbox}
\end{center}
\end{table*}

\end{document}